\documentclass[aps,prb,twocolumn,showpacs,superscriptaddress]{revtex4}
\usepackage{graphicx}

\begin{document}

% Use the \preprint command to place your local institutional report
% number in the upper righthand corner of the title page in preprint mode.
% Multiple \preprint commands are allowed.
% Use the 'preprintnumbers' class option to override journal defaults
% to display numbers if necessary
%\preprint{}
%Title of paper
\title{In-plane magnetic field effect on the neutron spin resonance in optimally doped  
FeSe$_{0.4}$Te$_{0.6}$ and BaFe$_{1.9}$Ni$_{0.1}$As$_{2}$ superconductors}

\author{Shiliang Li}
\email{slli@aphy.iphy.ac.cn}
\affiliation{Beijing National Laboratory for Condensed Matter Physics, Institute of Physics, Chinese Academy of Sciences, Beijing 100190, China
}
\author{Xingye Lu}
\affiliation{Beijing National Laboratory for Condensed Matter Physics, Institute of Physics, Chinese Academy of Sciences, Beijing 100190, China
}
\author{Meng Wang}
\affiliation{Beijing National Laboratory for Condensed Matter Physics, Institute of Physics, Chinese Academy of Sciences, Beijing 100190, China
}\affiliation{
Department of Physics and Astronomy, The University of Tennessee, Knoxville, Tennessee 37996-1200, USA
}
\author{Hui-qian Luo}
\affiliation{Beijing National Laboratory for Condensed Matter Physics, Institute of Physics, Chinese Academy of Sciences, Beijing 100190, China
}
\author{Miaoyin Wang}
\affiliation{
Department of Physics and Astronomy, The University of Tennessee, Knoxville, Tennessee 37996-1200, USA
}
\author{Chenglin Zhang}
\affiliation{
Department of Physics and Astronomy, The University of Tennessee, Knoxville, Tennessee 37996-1200, USA
}
\author{Enrico Faulhaber}
\affiliation{
Gemeinsame Forschergruppe HZB - TU Dresden,
Helmholtz-Zentrum Berlin f$\ddot{u}$r Materialien und Energie, D-14109 Berlin, Germany
}
\affiliation{
Forschungsneutronenquelle Heinz Maier-Leibnitz (FRM-II), TU M$\ddot{u}$nchen, D-85747 Garching, Germany}
\author{Louis-Pierre Regnault}
\affiliation{
Institut Nanosciences et Cryog$\acute{e}$nie, SPSMS-MDN, CEA-Grenoble, F-38054 Grenoble Cedex 9, France
}
\author{Deepak Singh}
\affiliation{NIST Center for Neutron Research, National Institute of Standards and Technology, Gaithersburg, Maryland 20899, USA
}
\author{Pengcheng Dai}
\email{pdai@utk.edu}
\affiliation{
Department of Physics and Astronomy, The University of Tennessee, Knoxville, Tennessee 37996-1200, USA
}
\affiliation{
Neutron Scattering Science Division, Oak Ridge National Laboratory, Oak Ridge, Tennessee 37831-6393, USA
}
\affiliation{Beijing National Laboratory for Condensed Matter Physics, Institute of Physics, Chinese Academy of Sciences, Beijing 100190, China
}
\begin{abstract}
We use inelastic neutron scattering to study the effect of an in-plane magnetic field 
on the magnetic resonance in optimally doped superconductors FeSe$_{0.4}$Te$_{0.6}$ ($T_c=14$ K) and BaFe$_{1.9}$Ni$_{0.1}$As$_{2}$ ($T_c=20$ K). While the magnetic field up to 14.5 Tesla does not change the energy of the resonance, it particially suppresses $T_c$ and the corresponding superconductivity-induced intensity gain of the mode. However, we find no direct evidence for the field-induced spin-1 Zeeman splitting of the resonance. Therefore, it is still unclear if the resonance is the long-sought singlet-triplet excitation directly coupled to the superconducting electron Cooper pairs.
\end{abstract}

% insert suggested PACS numbers in braces on next line


%\maketitle must follow title, authors, abstract, \pacs, and \keywords
\maketitle

The neutron spin resonance is arguably the most important collective magnetic excitations in unconventional superconductors that near an antiferromagnetic (AF) instability \cite{RossatMignodJ91,WilsonSD06,SatoNK01,ChristiansonAD08,LumsdenMD09,ChiS09,MookHA10,QiuY09}. Experimentally, the resonance can be broadly defined as a superconductivity-induced gain in the magnetic scattering and the corresponding imaginary part of the dynamical susceptibility, $\chi^{\prime\prime}(\omega,Q)$, at the AF wave vector.  The hallmark of the resonance is the increase of its 
magnetic intensity below $T_c$ like a superconducting order parameter \cite{RossatMignodJ91,WilsonSD06,SatoNK01,ChristiansonAD08,LumsdenMD09,ChiS09,MookHA10,QiuY09}.
Although the microscopic origin of the resonance is still unclear, the
mode is generally believed to arise from the spin-1 singlet-triplet excitations of the electron Cooper pairs \cite{EschrigM06}. Here, the process of the singlet-to-triplet excitation of an electron Cooper pair can be denoted as $\left|0\right\rangle \Rightarrow \left|1\right\rangle$ [Fig. 1(a)], where $\left|0\right\rangle$ = $\frac{1}{\sqrt{2}}(\left|\uparrow\downarrow\right\rangle - \left|\downarrow\uparrow\right\rangle)$ and $\left|1\right\rangle$ = $\left\{\left|\uparrow\uparrow\right\rangle, \frac{1}{\sqrt{2}}(\left|\uparrow\downarrow\right\rangle + \left|\downarrow\uparrow\right\rangle), \left|\downarrow\downarrow\right\rangle\right\}$ are singlet and triplet states, respectively. In some cases, the  resonance maybe a singlet-to-doublet excitation that only involves the $\left|\uparrow\uparrow\right\rangle$ and $\left|\downarrow\downarrow\right\rangle$ states [11] as shown in Fig. 1(c).

\begin{figure}
\includegraphics[scale=.45]{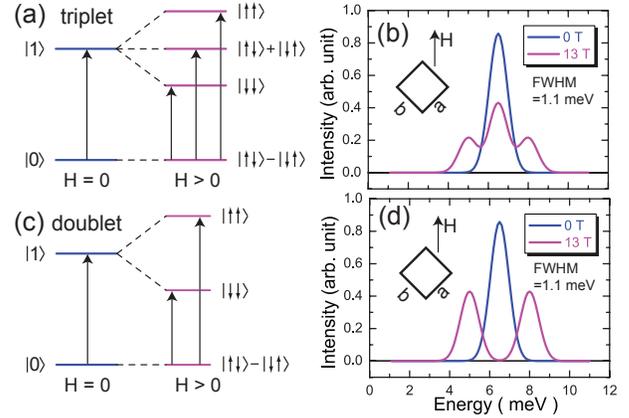}
\caption{ (Color online) (a) Schematic diagram of the Zeeman splitting of the exciton from singlet $\left|0\right\rangle$ to triplet $\left|1\right\rangle$ excited states. (b) Schematic diagram of the Zeeman splitting of a spin-1 resonance under 13-T field. The FWHM of the resonance is assumed to be close to the resolutions of our instruments estimated from phonon measurements. The intensity of the central peak is set to be equal to the sum of the other two peaks, which is a requirement of isotropic spin fluctuations. The Land$\acute{e}$ factor is taken as $g=2$. (c)-(d) Similar schematic diagrams as those in (a) and (b) in the case of a singlet-to-doublet excitation. 
 }
\end{figure}
 
If the resonance is indeed a spin-1 singlet-triplet excitation \cite{EschrigM06}, it should Zeeman split into three peaks under the influence of a magnetic field.  The degenerate triplet state $\left|1\right\rangle$ will split into three energy levels following the Zeeman energy $\Delta E=\pm g\mu_B B$ [Fig. 1(b)] \cite{LorenzoJE07}, where $g=2$ is the Lande factor
and $B$ is the magnitude of the field. If the resonance is the
 singlet-doublet excitation [Fig. 1(c)], 
the splitting of the mode under field can be observed as shown in Fig. 1(d).
For high-transition-temperature (high-$T_c$) copper oxide superconductors, application of a 14-T magnetic field only suppresses the intensity of the resonance with no evidence for the expected Zeeman splitting \cite{DaiP00,TranquadaJM04,WilsonSD07}.
In the case of Fe-based superconductors, where the neutron spin resonance is believed to arise from the electron-hole pocket excitations \cite{MazinII08} and has been found in hole/electron doped BaFe$_2$As$_2$ and
Fe(Te,Se) superconductors \cite{ChristiansonAD08,LumsdenMD09,ChiS09,MookHA10,QiuY09,ShamotoS10}, there  
are several neutron scattering experiments probing the effect of a magnetic field on the resonance.  In the first neutron measurement, 
application of a 14.5-T $c$-axis aligned magnetic field on the optimally electron-doped BaFe$_{1.9}$Ni$_{0.1}$As$_2$ 
($T_c=20$ K) was found to reduce the intensity
and shift down the energy of the mode with no evidence of the Zeeman splitting \cite{ZhaoJ10}. In separate experiments on the optimally superconducting Fe(Te,Se) \cite{LumsdenMD10,LiuTJ10,XuZ10,LeeSH10,ArgyriouDN10,LiS10}, a 7-T in-plane magnetic field can suppress the intensity of the resonance without shifting its energy or changing its width \cite{WenJ10}. For underdoped BaFe$_{1.92}$Ni$_{0.08}$As$_2$
($T_c=17$ K), where static AF order co-exists with superconductivity at zero field, application of a 10-T in-plane field enhances the AF order at the expense of the resonance, again with no evidence for the Zeeman splitting \cite{WangM11}.  Although a recent neutron scattering expriment on FeSe$_{0.4}$Te$_{0.6}$
using a $c$-axis aligned field suggests the presence of a field-induced Zeeman splitting of the resonance \cite{BaoW10}, it is still not clear whether the resonance is a spin-1 mode given the statistics of the data. Since a $c$-axis aligned field can suppress superconductivity much more efficiently \cite{LeiH10}, the best geometry to observe the Zeeman splitting is to align the magnetic field within the Fe plane, where the field-induced suppression of superconductivity is much less.                                                                                                   

In this paper, we report inelastic neutron scattering experiments measuring the effect of the in-plane magnetic field on the resonance in FeSe$_{0.4}$Te$_{0.6}$ ($T_c$ = 14 K, $\sim$ 4 grams, mosaic $\sim$ 2.5$^\circ$) and BaFe$_{1.9}$Ni$_{0.1}$As$_2$ ($T_c$ = 20 K, $\sim$ 6 grams, mosaic $\sim$ 2$^\circ$). Our experiments on FeSe$_{0.4}$Te$_{0.6}$  were carried out on PANDA cold neutron triple-axis spectrometer at Forschungsneutronenquelle
Heinz Maier-Leibnitz (FRM II), TU M$\rm\ddot{u}$nchen, Germany and on SPINS cold neutron triple-axis spectrometer at the NIST Center for Neutron Research (NCNR), USA.  We have also carried out in-plane field measurements on BaFe$_{1.9}$Ni$_{0.1}$As$_2$ using the IN22 thermal triple-axis
spectrometer at the Institut Laue-Langevin, Grenoble, France \cite{ZhaoJ10}.  For cold triple-axis measurements, we chose a fixed final neutron energy of $E_f=5$ meV and put a cooled Be filter before the analyzer. The energy resolution at elastic line is less than 0.25 meV. Pyrolytic graphite (PG) were used as the monochromator and analyzer.   We define the momentum transfer $Q$ at ($q_{x}$,$q_{y}$,$q_{z}$) as ($H$,$K$,$L$)=($q_{x}a/2\pi $,$q_{y}b/2\pi $,$q_{z}c/2\pi $) reciprocal lattice units (rlu), where
the lattice parameters of the tetragonal unit cell ($P4/nmm$ space group) are $a=b=3.786$ {\AA} and $c=6.061$ {\AA} for FeSe$_{0.4}$Te$_{0.6}$ and 
$a=b=3.963$ {\AA} and $c=12.77$ {\AA} for BaFe$_{1.9}$Ni$_{0.1}$As$_2$.  The samples were oriented in the 
$[H,H,L]$ scattering plane so that the applied field direction is along the $(1,-1,0)$ direction as shown in the inset of Fig. 1(b).
The setup for IN22 thermal thriple-axis measuremetns were described before \cite{ZhaoJ10}.

\begin{figure}[tbp]
\includegraphics[scale=.5]{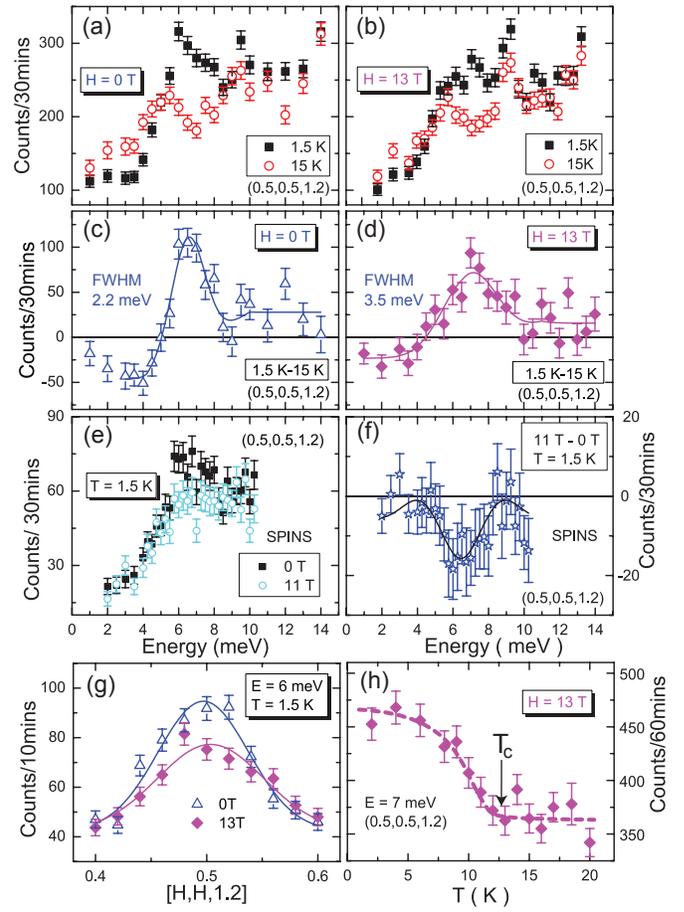}
\caption{(Color online) Constant-$Q$ scans at $Q=(0.5,0.5,1.2)$ at 1.5 K and 15 K under (a) 0 and (b) 13-T field. (c) and (d) show the magnetic resonance at 0 and 13-T, respectively, as determined by the difference of data between 1.5 K and 15 K. The solid lines are fitted results as described in the text. (e) Constant-$Q$ scans at (0.5,0.5,1.2) at 1.5 K under 0 and 11-T measured at SPINS. (f) The difference of data in (e) between 11 and 0-T. The solid line is fitted by the difference between one Gaussian and three Gaussians as illustrated in figure 1(b). (g) Constant-$E$ scans at 6 meV. (h) Temperature dependence of the resonance intensity at peak position ($E = 7$ meV) under 13-T field. 
  }
\end{figure}

We first describe the results on FeSe$_{0.4}$Te$_{0.6}$.  Figure 2(a) shows the zero-field constant-$Q$ scans at $Q$ = (0.5,0.5,1.2) at 1.5 K and 15 K. While the raw data show some features of unknown origin, the different spectra give rise to the effect induce by the superconductivity. 
The temperature difference spectrum in Fig. 2(c) shows a clear peak at $E=6.5\pm 0.2$ meV consistent with the previous results \cite{MookHA10,QiuY09,LiS10}. By fitting the peak with a Gaussian on a linear background as shown in Fig. 2(c), 
we find that the full-width-half-maximum (FHWM) of the resonance is $\Delta E=2.2\pm0.5$ meV.
Figure 2(b) shows identical constant-$Q$ scans at $Q= (0.5,0.5,1.2)$ taken under a 13-T in-plane magnetic field.  The temperature difference plot in Fig. 2(d) again shows a resonance that becomes broader in energy. A Gaussian fit to the data gives  
the peak position of $E=6.9\pm 0.4$ meV and the FWHM of $\Delta E= 3.5\pm 0.6$ meV [Fig. 2(d)]. 
Therefore, while a 13-T in-plane magnetic field broadens the resonance, there is no direct evidence for singlet-triplet splitting as suggested in Ref. \cite{BaoW10}. It is not clear whether such discrepancy comes from the different field directions applied in these two experiments. In principle, an in-plane magnetic field should offer better opportunity to observe field-induced Zeeman effect. We note, however, the FWHM of the resonance in Ref. \cite{BaoW10} (about 5 meV estimated from their data) is much larger than that in our experiment, which may suggest that our samples have better superconducting quality. 
A comparison of constant-$E$ scans at $E=6$ meV in zero and 13-T field  
demonstrates a clear suppression of the resonance intensity under the 13-T field [Fig. 2(g)].
Fig. 2(h) shows the temperature dependence of the scattering at $Q=(0.5,0.5,1.2)$ and 13 T, 
which increases below 12 K instead of 14 K in the zero field, consistent with the fact that the in-plane magnetic 
field also suppresses superconductivity.
To further demonstrate that the application of an in-plane magnetic field does not split the resonance, we have also carried out similar measurements on SPINS.
Fig. 2(e) shows the raw data taken at zero and 11-T field.  The field-on minus field-off data are shown in Fig. 2(f). Inspection of these data again reveal no evidence for Zeeman splitting. However, we note that 
the broadening of resonance in Fig. 2(d) is consistent with overlapping of three peaks 
with intrinsic FHWMs much larger than the instrumental resolution. No requirement of a large intrinsic anisotropic field is needed contrary to the suggestion in Ref \cite{BaoW10}. 
 
\begin{figure}[tbp]
\includegraphics[scale=0.5]{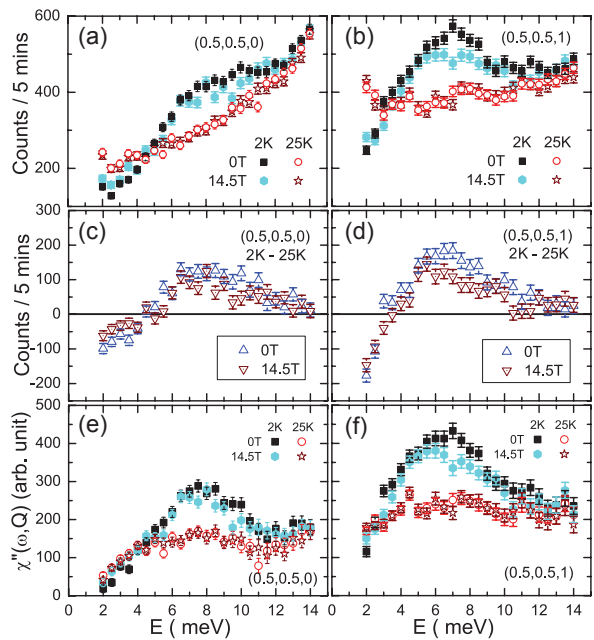}
\caption{ (Color online) Constant-$Q$ scans at (a) $Q=(0.5,0.5,0)$ and (b) $Q=(0.5,0.5,1)$. The corresponding temperature differences are shown in (c) and (d), while the $\chi^{\prime\prime}(Q,\omega)$ are shown in (e) and (f). }
\end{figure}

Having shown that there is no conclusive evidence for the Zeeman splitting of the resonance in FeSe$_{0.4}$Te$_{0.6}$, we now turn to our in-plane field measurements on BaFe$_{1.9}$Ni$_{0.1}$As$_2$. Figures 3(a) and 3(b) show constant-$Q$ scans at $Q=(0.5,0.5,0)$ and $Q=(0.5,0.5,1)$ below and above $T_c$
at zero and 14.5-T field \cite{WangM10}. In the normal state, a 14.5-T in-plane 
field has no observable effect on spin excitations at both wave vectors similar to that of a $c$-axis field \cite{ZhaoJ10}.  When cooling the system down to 1.5 K, the field again only has small effect on the spin excitations. Figures 3(c) and 3(d) show the temperature difference plots in zero and 14.5-T field.  Within the error of our measurements, a 14.5-T field has no observable effect to the resonance at $Q=(0.5,0.5,0)$ [Fig. 3(c)].  At wave vector $Q=(0.5,0.5,1)$, the field slightly suppresses the intensity of the resonance, again with no evidence for the expected Zeeman splitting. Figures 3(e) and 3(f) show 
our estimated imaginary part of the dynamic susceptibility, $\chi^{\prime\prime}(Q,\omega)$, at $Q=(0.5,0.5,0)$ and $Q=(0.5,0.5,1)$ below and above $T_c$ in zero and 14.5-T field respectively.  These results are obtained by subtracting the background and
correcting for the Bose population factor. They again suggest no observable field-induced effect at $Q=(0.5,0.5,0)$ and a small suppression of the resonance at $Q=(0.5,0.5,1)$.

\begin{figure}[tbp]
\includegraphics[scale=0.5]{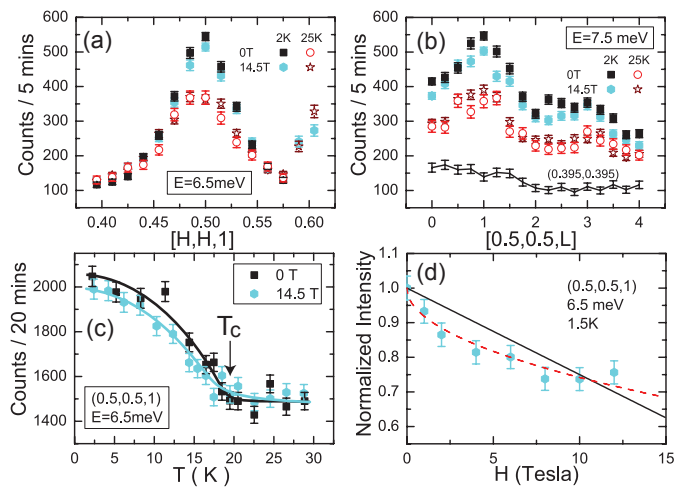}
\caption{ (Color online) (a) Constant-$E$ scans at 6.5 meV and $L = 1$. (b) $L$-scan at 7.5 meV (c) Temperature dependence of the resonance peak ($E = 6.5$ meV) intensity at (0.5,0.5,1). The lines are guided by eye. (d) Field dependence of the resonance peak intensity at (0.5,0.5,1). The solid and dash lines are fitted by the linear and square root functions respectively.
 }
\end{figure}

Fig. 4(a) gives constant-energy scans below
and above $T_c$ in zero and 14.5-T fields along the $[H,H,1]$ direction, 
the applied field only has limited effect on spin excitations in the superconducting 
state and no effect in the normal state.  Fig. 4(b) shows scans long the 
$[0.5,0.5,L]$ direction which again reveal weak magnetic field effect at 2 K.
Fig. 4(c) plots the temperature dependence of the scattering at the resonance energy.  In contrast to the 
earlier measurements for a $c$-axis aligned field \cite{ZhaoJ10}, the in-plane field has virtually no effect on
$T_c$ and only suppresses the intensity of the resonance moderately.
Fig. 4(d) shows the magnetic field dependence of the scattering at the resonance energy and $Q=(0.5,0.5,1)$.
In Ginzburg-Landau theory, the magnetic field dependence of the superconducting gap $\Delta(B)$ is related to the zero field
gap $\Delta(0)$ via $\Delta(B)/\Delta(0)=\sqrt{1-B/B_{c2}}$, where $B_{c2}$ is the upper critical field \cite{ZhaoJ10}.
Assuming that the intensity of the resonance is associated with the superconducting volume fraction or superfluid density, one would expect that 
the intensity of the mode to decrease linearly with increasing field \cite{DaiP00}, or $I/I_0=1-B/B_{c2}$.  We have used both linear (solid line) 
and square root (dash line), where $I/I_0=1-(B/B_{c2})^{1/2}$, relations to fit the field dependence data.  The outcome gives the $B_{c2}$ as 40-T and 150-T, respectively.  
While these results suggest that the square root relationship fits the data better, its physical significance is unclear.

In summary, we have studied the effect of an in-plane magnetic field on the neutron spin resonance of 
FeSe$_{0.4}$Te$_{0.6}$ and BaFe$_{1.9}$Ni$_{0.1}$As$_2$ superconductors. While our initial purpose is to study the Zeeman splitting
of the spin-1 triplet of the resonance, we are unable to conclusively establish that the mode is indeed a singlet-triplet excitation.
From recent polarized neutron scattering measurements on BaFe$_{1.9}$Ni$_{0.1}$As$_2$ \cite{LipscombeOJ10}, we know that the resonance is inconsistent with a simple isotropic singlet-triplet excitation, and appears only for spin moment parallel to the Fe plane, which may results in a singlet-doublet excitation [Fig. 1(c)]. This is consistent with the present magnetic field effect, where no field-induced broadening of the mode was observed since the picture in Fig. 1(d) is only applied to the isotropic case.  On the other hand, polarized neutron scattering experiments have found quasi-isotropic spin resonance
in FeSe$_{0.5}$Te$_{0.5}$ \cite{Babkevich10}, this may explain why an in-plane magnetic of 14.5-T can clearly broaden the resonance (Fig. 2).  Although these results may be consistent with the mode being a singlet-triplet excitation in FeSe$_{0.5}$Te$_{0.5}$, we cannot conclusively establish this based on the present data.  Regardless whether the resonance is a spin-1 singlet-to-triplet excitation, it is directly associated with superfluid density and superconducting volume fraction.  Therefore, understanding its microscopic origin is still important for determining the role of spin excitatins for high-$T_c$ superconductivity. 

This work is supported by Chinese Academy of Science, 973 Program (2010CB833102,
2010CB923002, 2011CBA00110), and by the
US DOE, BES, through DOE DE-FG02-05ER46202 and Division of Scientific User
Facilities.

\end{document}